\documentclass[useAMS,usenatbib]{mn2e}
\usepackage{graphicx}
\usepackage{amsmath}
\usepackage{amssymb}
\usepackage{color}
\usepackage{url}

\newcommand\nodata{ ~$\cdots$~ }%

\newcommand  \acc     {\ifmmode {\rm km\,s}^{-2} \else km\,s$^{-2}$\fi}

\newcommand  \ergs     {\ifmmode {\rm ergs\,s}^{-1} \else ergs s$^{-1}$\fi}
\newcommand  \ergcms   {\ifmmode {\rm erg~cm}^{-2}\,{\rm s}^{-1}
                        \else erg~cm$^{-2}$\,s$^{-1}$\fi}
\newcommand  \ergcmsA  {\ifmmode{\rm erg\,cm}^{-2}\,{\rm s}^{-1}\,{\rm\AA}^{-1}
                        \else ergs\,cm$^{-2}$\,s$^{-1}$\,\AA$^{-1}$\fi}
\newcommand  \ergcmsHz {\ifmmode{\rm ergs\,cm}^{-2}\,{\rm s}^{-1}\,{\rm Hz}^{-1}
                        \else ergs\,cm$^{-2}$\,s$^{-1}$\,Hz$^{-1}$\fi}
\newcommand  \phcms    {\ifmmode {\rm ph\,cm}^{-2}\,{\rm s}^{-1}
                        \else ph\,cm$^{-2}$\,s$^{-1}$\fi}
\newcommand  \phcmsA   {\ifmmode {\rm ph\,cm}^{-2}\,{\rm s}^{-1}\,{\rm\AA}^{-1}
                        \else ph\,cm$^{-2}$\,s$^{-1}$\,\AA$^{-1}$\fi}
\newcommand\actaa{{Acta Astronomica}}% 
\newcommand\aj{{AJ}}% 
\newcommand\araa{{ARA\&A}}% 
\newcommand\apj{{ApJ}}% 
\newcommand\apjl{{ApJ}}% 
\newcommand\apjs{{ApJS}}% 
\newcommand\aap{{A\&A}}% 
\newcommand\aapr{{A\&A~Rev.}}% 
\newcommand\aaps{{A\&AS}}% 
\newcommand\mnras{{MNRAS}}% 
\newcommand\prd{{Phys.~Rev.~D}}% 
\newcommand\pasp{{PASP}}% 
\newcommand\pasj{{PASJ}}% 
\newcommand\solphys{{Sol.~Phys.}}% 
\title[The case for nearby FRBs]
{Fast radio bursts: the observational case for a Galactic origin}
\author[D. Maoz et al.]
{Dan Maoz$^{1}$, Abraham Loeb$^{1,2}$, 
Yossi Shvartzvald$^{1,3}$, Monika Sitek$^{4}$,\cr 
Michael Engel$^{1}$,
Flavien Kiefer$^{1}$, 
 Marcin Kiraga$^{4}$,Amir Levi$^{1}$,
Tsevi Mazeh$^{1}$,\cr
Micha{\l} Pawlak$^{4}$,
R. Michael Rich$^{5}$,
Lev Tal-Or$^{1}$, 
Lukasz Wyrzykowski$^{4}$\\ 
{\small{\it $^{1}$School of Physics and Astronomy, Tel-Aviv University, Tel-Aviv 69978,
Israel}}\cr
{\small {\it $^{2}$Institute for Theory and Computation, 
Harvard University, Cambridge, MA 03210, USA}}\cr
{\small {\it $^{3}$NASA Postdoctoral Program Fellow, Jet Propulsion Laboratory, Pasadena, CA 91109, USA}}\cr
{\small {\it $^{4}$Warsaw University Observatory, 00478 Warsaw, Poland}}\cr
{\small {\it $^{5}$Department of Physics and Astronomy, University of California, Los Angeles, CA 90095, USA}}
}\date{\today}

\begin{document}

\maketitle

\label{firstpage}

\begin{abstract}
There are by now ten published detections of 
fast radio bursts (FRBs)---single bright GHz-band millisecond pulses of unknown 
origin. Proposed explanations cover a broad range from exotic processes at 
cosmological distances to atmospheric and terrestrial sources. 
Loeb, Maoz, and Shvartzvald have previously suggested
that FRB sources could be nearby flare stars, and pointed out the presence 
of a W-UMa-type contact binary within the beam of one out of three FRB
fields that they examined. To further test the flare-star hypothesis,
we use time-domain optical photometry and spectroscopy,
and now find possible flare stars in additional FRB fields, with one 
to three such cases among all eight FRB fields studied.
We evaluate the chance probabilities of these possible associations
to be in the range $\sim 0.1\%$ to 9\%, depending on the input assumptions.
%The random probability 
%for  having four (or one) real W UMa's 
%within FRB beam areas is $\sim 10^{-5}$ (or 0.14\%--9\%).
Further, we re-analyze the probability 
that two FRBs recently discovered 3 years apart 
within the same radio beam are unrelated.
Contrary to other claims, we conclude with $99\%$ confidence
that the two events are from the 
same repeating source. The different dispersion measures between the two bursts 
then rule out a cosmological intergalactic-medium origin 
for the dispersion measure, but are consistent with the flare-star 
scenario with a varying plasma blanket between bursts. Finally,
we review some theoretical objections that have been raised against a 
local flare-star FRB origin, and show that they are incorrect.   
\end{abstract}


%.

%\newpage

\newpage

\section{Introduction}

Nearly a decade after their discovery by \citet{lorimer07}, the 
source and nature of
fast radio bursts
(FRBs) is yet unknown.
FRBs are
bright ($\sim 0.1-1~{\rm Jy}$) and brief ($\sim 1~{\rm ms}$)
pulses of $\sim 1~{\rm GHz}$ radio emission. The latest
estimate of the rate of FRBs with flux $>0.1$~Jy is $3.3^{+2}_{-1}\times
 10^3~{\rm day}^{-1}$ over the whole sky
\citep{rane15}. To date, ten candidate bursts have been reported 
\citep{lorimer07,keane12,thornton13,spitler14,burke-spolaor14,petroff15a,
ravi15}.
Nine of them have 
been discovered with the Parkes Radio Telescope, and one with the Arecibo
Radio Telescope. Eight of the FRBs were found in the analysis of archival data,
long after the events. Two FRBs were discovered in real time (within 10~s)
\citep{ravi15,petroff15a}, one of them while observing the Carina dwarf galaxy,
the other
while re-monitoring the location of a previous burst (FRB110220, 
\citet{thornton13}). The latter 
 was also the first FRB with a polarization 
measurement, and was found to be up to $\sim 50\%$ 
circularly polarised, depending 
on the temporal part of the pulse considered.
In no cases have there been known transient or eruptive counterparts  
at other wavelengths coincident with the FRB, including in the real-time
detection by \citet{petroff15a}, 
where prompt followup observations at other bands were undertaken 
(within 9 hrs in X-rays and within 
17 hrs in the optical and near infrared).  
No clear consensus has
emerged on whether FRBs are isotropically distributed, or avoid
the Galactic plane, due to some combination of backgrounds, selection effects, 
and propagation
effects \cite[see, e.g.][]{petroff15a,rane15,macquart15}. An intrinsic 
avoidance of the Galactic plane is not expected in any model that has
been proposed for FRBs.  

The dispersion measures
(DMs) of FRBs, i.e. the line-of-sight column densities of free electrons,
as deduced from the normalisation of the 
$f^{-2}$ frequency dependence of the pulse arrival times,  
exceed the values expected from models of the Milky Way interstellar 
electron distribution. FRBs have  DMs in the range $\sim 
300-1600$~pc~cm$^{-3}$, or electron 
column densities $N_e\sim (1-5)\times 10^{21}~{\rm cm}^{-2}$.
FRBs have therefore been inferred to
originate from extragalactic sources at cosmological distances.
 The intergalactic medium, perhaps combined with the DM of the sources'
host galaxies, would account for the excess DM values. In one case, however
(FRB010621, \citet{keane12}), direct estimates of the ionized gas column 
density in this direction show that the Galactic models underestimate the 
DM, placing the FRB source within the Galaxy at a distance 
of 8-20~kpc \citep{bannister14}, or possibly much closer if some of the DM
is intrinsic to the source.
In addition to the excess DM, three or four of the FRBs 
(\citet{lorimer07}, one from \citet{thornton13}, \citet{ravi15}, and 
possibly \citet{burke-spolaor14}) 
show frequency-dependent pulse 
smearing, indicative of scattering by an inhomogeneous electron distribution
 along the
line of sight, at a level much larger than expected by the Galactic
interstellar medium. The intergalactic medium has been invoked to
explain this scattering as well (e.g. \citet{macquart13}), 
but \citet{katz14,katz15} 
has argued that the implied densities are too high and the scattering 
instead likely occurs near the source.  

If FRB sources are extragalactic, their bright fluxes imply 
high isotropic luminosities of $L_{\rm iso}\gtrsim 10^{42}$~erg~s$^{-1}$.
A flurry of theoretical ideas has been put forward to explain 
them, including: black-hole evaporation \citep{keane12}; magnetar hyperflares \citep{popov13,katz14}; 
neutron-star mergers \citep{totani13,ravi14}; white-dwarf mergers \citep{kashiyama13}; 
collapse of supramassive 
neutron stars \citep{falcke14,zhang14}; orbiting bodies immersed in pulsar winds \citep{mottez14}; 
magnetar pulse-wind interactions \citep{lyubarsky14}; giant pulses from magnetars
near galactic centers \citep{pen15};  collisions between 
neutron stars and asteroids \citep{geng15}; giant pulses from 
young pulsars within supernova remnants in nearby galaxies \citep{connor15} or 
at cosmological distances \citep{katz15}; quark novae \citep{shand15};
and even directed signals from extraterrestrial civilizations
\citep{luan14}. 

In parallel, the possibility has lingered that FRBs have
a terrestrial, possibly human-made, origin \citep{burke-spolaor11,kulkarni14}. From the start,
some resemblances were
 noted between FRBs and ``perytons'' \citep{burke-spolaor11}, 
signals
with fluxes, time structures, and roughly quadratic frequency sweeps
 similar to FRBs but, as 
opposed to FRBs, with detections in all focal-plane receivers of the telecope, 
indicating a nearby source that has not been focused by the antenna dish.
\citet{petroff15b} have recently demonstrated that perytons arise from 
microwave kitchen ovens at the Parkes observatory (generally during lunchtime), 
when an oven door is opened while the appliance is in operation. They 
presented further statistical evidence that the FRBs are a truly astronomical 
phenomenon, distinct from the human-made perytons.      
 
Alternatively to these explanations for FRBs, \citet*{loeb14l} proposed 
that FRB sources could be nearby flaring stars. 
Abundant types of flare stars that might provide the required FRB rates
include young M-type stars, W-UMa-class stars (stable contact binaries composed
of two stars of about solar mass), or other types. Flaring 
main-sequence stars are already known to produce coherent radio
bursts with millisecond-scale durations and $\sim 1$~GHz
flux densities of a fraction of a Jy \citep{lang83,lang86,guedel89,bastian90}. 
Although details of the flare
emission mechanisms are poorly understood, these bursts are
thought to be produced by a cyclotron maser process, in which
bunches of electrons emit coherent radio emission 
as they gyrate around the magnetic
field of the host star \citep{guedel02,treumann06,matthews13}.
\citet{loeb14} argued that 
if FRB eruptions are produced at the bottom of the coronae of their
host stars, or the eruptions 
cause coronal mass ejections of their own \citep{drake13}, 
the radiation will pass through a
plasma blanket with a characteristic electron column density $\gtrsim
10^{10}~{\rm cm^{-3}}\times R_\odot\sim 300~{\rm cm^{-3}~pc}$, as
needed to explain the observed DMs of   FRBs. Such column densities are typically inferred for 
stellar flares \citep{getman08a, getman08b}
We note that radio bursts from 
stellar flares are observed to be circularly polarized, like the single 
FRB with a polarization measurement \citep{petroff15a}.

To test the flare-star FRB hypothesis, \citet{loeb14} monitored in the optical 
the 
fields of three FRBs for variable stars. In one field, they discovered a nearby
(800~pc) W UMa system at a position within the FRB radio beam. From the 
abundance of W UMa systems, \citet{loeb14} derived a probability of 
$\sim 5\%$ for a chance coincidence of such a system in the beam of one out of 
three FRB beams. However, this inference was compromised by the 
{\it a posteriori} nature of the conclusion, since the objective of 
specifically finding W UMa stars was not defined in advance of the search.

In this paper, we extend the observational test of the flare-star hypothesis,
by means of photometric and spectroscopic monitoring and analysis
 of stars in 8 out of 9 known FRB fields. We
find one new example of a W UMa stars in the direction of one FRB, and
a periodically variable bright blue star toward another FRB. 
We also revisit the statistical analysis of FRB110220 and FRB140514 and conclude
that they are from one and the same repeating source, rather than 
from two unrelated sources as claimed by \citet{petroff15a}. 
The factor-of-2 decrease over 3 years in DM between the two bursts cannot 
arise in the intergalactic medium,
invalidating any extragalactic model for FRBs in which the DM is a measure of 
the distance to the source. Finally,   
we re-assess some
arguments that have been put forward against the Galactic flare-star scenario,
and show that they are incorrect from a theoretical perspective,
from a demonstrative observational point of view, or both.
  
\section{Optical observations and analysis of FRB fields}

Following the methodology of \citet{loeb14} to search for bright
variable stars within the radio beams of past FRBs, we have analyzed
data at those locations from the All-Sky Automated Survey (ASAS) 
\citep{pojmanski97}. ASAS uses a set of small telecopes in Chile 
to monitor about 3/4 of the sky to 14~mag in the optical ($V$ and $I$), 
with data for some fields going back two decades. 
The ASAS catalog also permits us to find the  
sky density of specific types of variable stars
 at the Galactic longitude and latitude, $(l,b)$
of each FRB, and thus to estimate the chance probability of finding 
such a star within an FRB beam.

\subsection{FRB110703} In the field of this FRB, found by 
\citet{thornton13}, 
\citet{loeb14} discovered, based on imaging with 
the Wise Observatory 1m telescope, a $V=13.6$~mag, $V-I=0.85$~mag, 
W UMa system with a period of 
$P=7.84$~hr within the 14~arcmin diameter Parkes beam,  4~arcmin 
 from the beam center. 
 From the number (one) of W~UMa's that are as bright which they found 
outside the beam region,
and also from the known volume 
density of W UMa's,  \citet{loeb14} roughly estimated a $\sim 1.7\%$ 
chance 
probability for finding such a system within the FRB beam region. ASAS data 
for this field confirm the $V$ and $I$ light-curve shapes, amplitudes, 
and period for this W UMa system (named ASAS233003-0248.3 
in the ASAS database). Furthermore, from the ASAS catalog, we find that the 
actual sky density of W UMa's in the general direction 
of this FRB, to $V=14$~mag, is 0.11~deg$^{-2}$, implying only a 0.5\% 
random probability to find a W UMa within the 0.04~deg$^{-2}$ Parkes beam
(i.e. lower than the rough estimate, above, by \cite{loeb14}.) 

\begin{figure}
 \centering
 \includegraphics[width=6cm, angle=-90
]{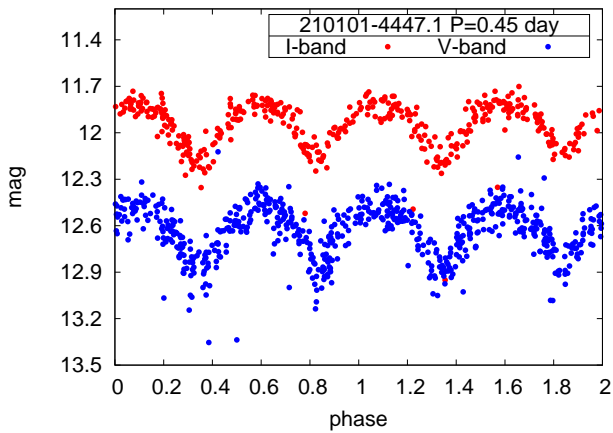}
\caption{Phase-folded $V$ and $I$ light curves from ASAS for the variable
star, ASAS220101-4447.1, a possible W UMa-type source for FRB110627.}
\label{fig1}
\end{figure}

\subsection{FRB110627} 
Near the position of 
FRB110627 \citep{thornton13}, we have found another bright ($V=12.5$ mag)
W UMa system, 
ASAS220101-4447.1. Figure~\ref{fig1} shows the 
$V$ and $I$ light curves, folded with a period of $P=10.75$~hr. The period,
the color ($V-I=0.7$), and the light-curve shape (with amplitude 0.4~mag)
are all characteristic of a W UMa star. 
The star is 29 arcmin west, i.e. about 2 beam diameters, from the center of  
beam No. 12 in which the FRB was detected\citet{keane15}.
However, beam No. 12
is on the outer ring of beams of the 
Parkes multi-beam receiver\footnote{see e.g., {\tt http://www.atnf.csiro.au/research/multibeam/instrument/description.html}}. 
 From the technical data for FRBs at the 
Reasearch Data Australia Portal\footnote{\tt http://http://supercomputing.swin.edu.au/data-sharing-cluster/parkes-frbs-archival-data/}, 
the receiver orientation on the sky 
at the time of the observation was such that beam No. 12 was almost exactly 
on the western side of the outer ring (i.e. at the ``3 o'clock'' position).  
%at the time and position of the 
%FRB, the telescope was pointed due west and 
As such, the detection in this single beam could have been triggered
by a side lobe of a bright FRB at the west-offset position of the W UMa. 
The FRB/W-UMa association is thus possible in this case. From ASAS, the 
W UMa sky density in this FRB's direction is 0.16~deg$^{-2}$. 
Considering that 
the area within which this type of side-lobe detection in 
beam No. 12 could have been triggered (an annular sector of opening angle $60^{\circ}$, centred on the receiver centre, 
with inner radius 65 arcmin 
and outer radius 80 arcmin, i.e. at the location of the W UMa star) is about 0.3~deg$^{2}$, 
the chance probability for the presence of a W UMa there is about 5\%.  

\begin{figure}
 \centering
 \includegraphics[
width=6cm, angle=-90
]{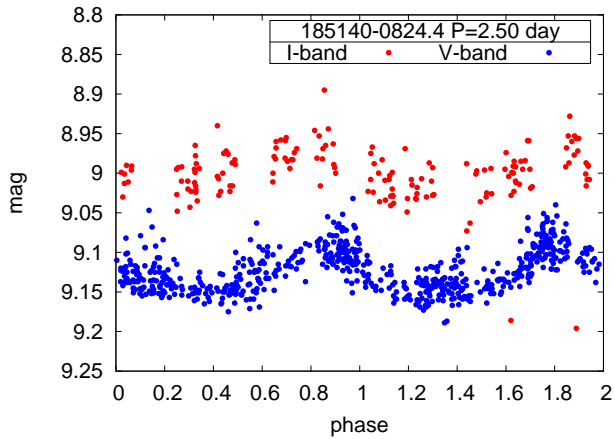}
\caption{Phase-folded $V$ and $I$ light curves from ASAS for the bright variable
blue star, HD174628, within the beam of FRB010621.}
\label{fig2}
\end{figure}
  
\subsection{FRB010621} In the field of this FRB discovered by \citep{keane12}, 
ASAS data \cite[Fig.~\ref{fig2}, see]{pojmanski05} 
show that the bright blue star ($V=9.1$~mag, $V-I=0.1$), HD174628,
6.7~arcmin from the reported FRB radio beam center, is variable with
 a period of 2.5~days and a $V$-band
amplitude of $\sim 0.07$~mag. 
Figure~\ref{HD17spectrum} shows the star's spectrum (see below) 
with its best-fitting 
model template, a B9~III giant with effective temperature $T\sim 11,000 K$,
surface gravity $\log g=3.5\pm 0.5$, and projected rotational broadening 
$V_{\rm rot}\sin i=70\pm 10$~km~s$^{-1}$.
\begin{figure}
 \centering
 \includegraphics[
width=9.5cm
]{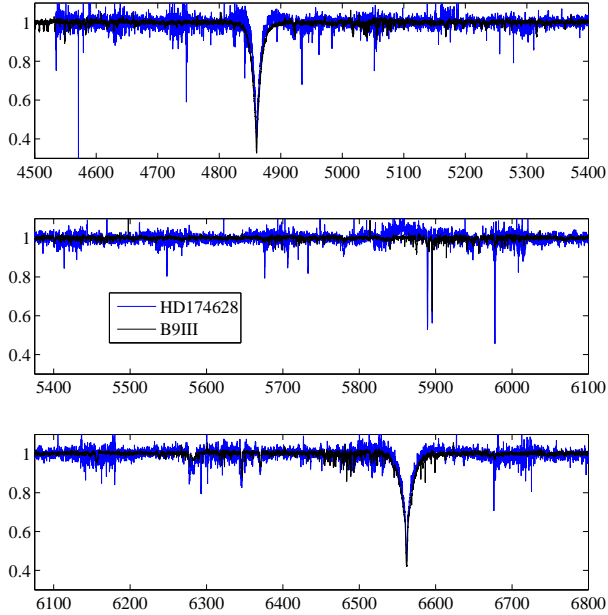}
\caption{Continuum-normalized Wise eShel spectrum of the 
periodic variable HD174628, with its best-fitting 
model stellar template, a B9~III star. H$\alpha$ and H$\beta$ are the prominent lines.}
\label{HD17spectrum}
\end{figure}
The spectral type, period, and rotational broadening are all characteristic
of a class of non-radially  
pulsating stars called slowly pulsating B-stars \cite[SPBs; e.g.][]{decat07}.
HD174628 is somewhat unusual, however, in that 
SPBs typically have  pulse amplitudes of only 0.01, and 0.025 at most. 
%(although the period could be double this value, i.e.
%2.3~days, if e.g. every
%period includes 
%two perspectives of an ellipsoidally-distorted star). 

To test for radial-velocity variations, or for the possibility that the 
photometric variations are caused by some other star along the line of sight,
whether or not physically associated with the B star, 
we obtained high-resolution spectra at 8 separate
epochs, as follows. Spectra with resolution 
$R\approx 10,000$ were obtained using the Wise Observatory 1m telescope 
with the eShel high resolution fiber-fed echelle 
spectrograph on four nights over a 32-day 
period, as detailed in Table~\ref{specjournal}. A one-hour exposure was
taken each night. A month later, the Levy echelle spectrometer on the Automated Planet 
Finder \cite[APF;]{vogt14} 2.4m telescope at Lick Observatory was used to further obtain 
$R\approx 100,000$ spectra at three epochs over 100 days. Each epoch consisted
of four consecutive 15-min exposures that were analyzed individually. Both data
sets were reduced and calibrated using standard procedures. 

We have processed each individual spectrum through the TODCOR two-dimensional 
correlation program \citep{zucker94}, attempting to fit each epoch with 
two stellar templates with variable radial velocities (RVs). In no cases do 
we find evidence of a secondary, late-type stellar component in the spectra.
The spectral features in the spectrum consist of only
 a handful of broad absorption lines, and 
therefore the RV precision is not much better than $c/R$. In the Wise data,
we estimate a single-epoch RV  precision of $\sim 4$~km~s$^{-1}$. In the 
Lick-APF spectra, from the scatter in RV among each set of four individual 
exposures taken over an hour, we estimate the RV precision at 
$\sim 1$~km~s$^{-1}$. As listed in Table~\ref{specjournal}, 
on all but one of the 8 nights that the star
was observed, the RV is consistent with a constant barycenter-corrected
 value of 
$\approx -16$~km~s$^{-1}$. One APF epoch shows a formally significant 
deviation to an RV of 
$\approx -22$~km~s$^{-1}$, but given the consistent velocities at all other 
epochs, at various phases of the 2.5~day photometric period, we suspect this
single deviant point is a fluke.      
We detect also no significant differences in $V_{\rm rot}\sin i$ among epochs. 

We have thus found no explanation for the 7\%-level, few-day-period
 photometric modulation 
of this star, other than it being an SPB star with an unusually high amplitude.
A variable line-of sight 
late-type star responsible for the variations would have contributed at least
 10\% of the B-star flux, which would have been detected in the spectrum
and can thus be ruled out. A contact binary composed, e.g.,
of two $10 M_\odot$ B-type giants with a separation of $25 R_\odot$ would 
have an orbital period of 3.3~days. If viewed nearly pole-on to the orbital 
plane, the photometric  variability at the $\pm 5\%$ level could be
due to ellipsoidal tidal distortion of the components. However, 
to satisfy the constraint
of RV  variations with amplitude $K<2$~km~s$^{-1}$, despite the 200~km~s$^{-1}$
orbital velocity, the system would need to be within 0.5 degree of pole-on. 
Furthermore, one would then expect the synchronized 
projected rotation velocity to be equally small, rather than the observed  
$v\sin i=70\pm 10$~km~s$^{-1}$. The same problems would arise, but would be
less severe, if the binary
companion in the low-inclination orbit were a roughly solar-mass 
object---a main-sequence star, white dwarf, neutron star or black hole.
For periods of $3.3$~days, the B-star's orbital velocity would be of order 
30-40~km~s$^{-1}$, and thus an inclination within 5 degrees of pole-on
would be required, in order to satisfy the observed $\Delta$RV$\lesssim 4$~km~s$^{-1}$ 
limits.
About one-half of SPB stars are magnetic, with fields of 
order hundreds of Gauss and sometimes close to $\sim 1$~kG \citep{hubrig09}, 
not 
unlike the $\sim 1$~kG fields of low-mass flare stars. B stars are relatively
rare, and more so SPB stars. For example, in the 105 deg$^2$ {\it Kepler} field, there are only 
about 10 B stars brighter than 9~mag, and only 4 of them are SPBs (all with amplitudes 1-3 orders
of magnitude smaller than HD174628). 
%Similarly, within 100 deg$^2$
%around HD174682 there in the ASAS database no other variable stars with 
%$V-I<0.1$. 
 A density of $<0.04$~deg$^{-2}$
\citep{mcnamara12}, can thus serve as a firm upper limit for SPBs
with the large HD174628 pulsation amplitude.   
The presence of this unusual star in the 0.042 deg$^{2}$ Parkes 
beam in one out of 8 FRB fields studied (random probability $<1.4\%$)
is therefore intriguing. However, again, care must be taken 
with such {\it a posteriori} statistical conclusions, and  therefore
we cannot claim a secure association between the star and the FRB. 
 
We note again that, for this FRB, \citet{bannister14} have used velocity-resolved observations
of H$\alpha$ and H$\beta$ Balmer line emission from diffuse ionized gas  
in the FRB direction, combined with a Galactic rotation model, to obtain the emission
measure of the Milky Way ionized gas column along the line of sight. From the emission
measure, they estimated the Galactic electron column density. They concluded that 
the DM of the FRB, if all due to the Galactic interstellar medium, 
places the FRB source within our Galaxy, at a distance 
of 8-20~kpc. The B star HD174628, discussed above, is at a distance of $0.4-1$~kpc,
depending on its exact spectral class and the line-of-sight 
extinction that it undergoes.

\begin{table*}
  \caption{HD174628 spectral observations and radial velocities}
  \label{specjournal}
  \centering
 \begin{tabular}{p{1.6cm}p{1cm}p{1.cm}p{.8cm}p{1.8cm}}
  \hline
 UT& UT &  JD& Observ.& RV \\
Date& Time&$-$2456700&&[km~s$^{-1}$]\\
\hline
2014.05.14 &01:12& 91.55 &Wise& $ -14.4\pm 3.4$\\ 
2014.06.10 &23:02&119.46 &    & $ -17.5\pm 2.7$ \\
2014.06.12 &00:58&120.54 &    & $ -14.0\pm 4.0$ \\
2014.06.14 &00:43&122.53 &    & $  -8.3\pm 4.3$ \\
2014.06.14 &21:07&123.38 &    & $ -12.5\pm 3.9$ \\
2014.07.10 &06:43&148.78 &Lick& $ -21.5\pm 1.1$\\
2014.10.14 &02:53&244.62 &    & $ -15.3\pm 1.0$\\
2014.10.18 &03:07&248.63 &    & $ -16.8\pm 0.9$\\

\hline
\end{tabular}
\end{table*}

\begin{table*}
  \caption{FRBs, candidate counterparts, and random probabilities for W UMa}
  \label{frbtable}
  \centering
 \begin{tabular}{p{1cm}p{.4cm}p{.4cm}p{3cm}p{1.1cm}p{1.3cm}p{
1.1cm}
p{1.3cm}p{.8cm}p{.5cm}p{4.4cm}}
  \hline
      &    & FRB&                   &          &         & Counterpart&      &$\Sigma$&Prob& \\
Date   &  l &  b & Ref.              & RA(J2000)&DEC      & RA(J2000)& DEC      & deg$^{-2}$& \%  &Comments\\
(1)&(2)&(3)&(4)&(5)&(6)&(7)&(8)&(9)&(10)&(11)\\
\hline
010621&  25& -04&\citet{keane12}        &18 52    &-08 29    & 18 51 40 &-08 24.4  &   0.16     &4.4 & B giant,$V=9.1$~mag, beam 10\\
010824& 301& -42&\citet{lorimer07}      &01 18 06 &-75 12 19 & \nodata  &  \nodata &   0.50$^*$ &0.13& \nodata\\
011025& 357& -20&\citet{burke-spolaor14}&19 07    &-40 37    & \nodata  &  \nodata &   0.19&0.8 &  \nodata\\
110220&  51& -55&\citet{thornton13}     &22 34    &-12 24    & \nodata  &  \nodata &   0.10&0.5 &\nodata\\
110627& 356& -42&\citet{thornton13}     &21 03 44 &-44 44 19 & 21 01 01 &-44 47.0  &   0.16&5.0 & W UMa, $V=12.4$~mag, beam 12\\
110703&  81& -59&\citet{thornton13}     &23 30    &-02 52    & 23 30 03 &-02 48.3  &   0.10&0.5 & W UMa, $V=13.5$~mag\\
120127&  49& -66&\citet{thornton13}     &23 15    &-18 25    & \nodata  &   \nodata&   0.11&0.5 & \nodata \\      
121102& 175& -0 &\citet{spitler14}      &05 32 10 & 33 05 13 &~~~~~~~~~~not&  searched&   0.15&--  & Arecibo FRB, Galactic plane   \\ 
131104& 261& -22&\citet{ravi15}         &06 44 10 &-51 16 40 & \nodata  &   \nodata&   0.18&0.8 & \nodata     \\  
140514&  51& -55&\citet{petroff15a}     &22 34 06 &-12 18 47 & \nodata  &   \nodata&   --  &--  & Site of FRB110220\\
      \hline
     \\
\end{tabular}
{Notes.- (2)(3)--Galactic longitude and latitude; (7)(8) coordinates of candidate stellar counterpart; (9) sky density,
to $V=14$~mag, of W UMa stars; (10) Probability of finding a W UMa within the FRB beam area. $^*$ density of W UMa's to $V=18.5$~mag,
in the SMC field.} 
\end{table*}

\subsection{Non-detections} Apart from the above three FRB fields, we 
have searched another five FRB fields for bright short-period 
($\lesssim 1$~day) variable stars,
but have not found any, as follows. The fields of 
FRB110220 and FRB120127 were already monitored 
and reported by \citet{loeb14}, who found no short-period stars
brighter than 16~mag. This null 
result is confirmed by the shallower but longer-term ASAS data for these two fields,
as well as for FRB011025 \citep{burke-spolaor14}. 
Near the position of FRB131104 \citep{ravi15}, ASAS data reveal a W UMa with 
$V=13.5$, $V-I=1$, and a period of $P=7.6$~hr. However, the star is 28~arcmin
south of the center of beam No. 5 in which the FRB was detected.
Given the receiver orientation at the time \citep{ravi15},
an FRB from the W UMa position would have triggered beam No. 12, rather than
No. 5. This W UMa star is therefore not the source of this FRB.

FRB010824, the first \citep{lorimer07} FRB discovered, 
adjacent to the Small Magellanic 
Cloud (SMC), and at 30~Jy still the brightest one known,
is also the only one detected in more than one focal-plane beam of the Parkes
multi-beam receiver. 
The three beams with detections permit a more precise localization of the 
FRB (see \cite{kulkarni14}, and also \cite{keane15}, 
who report a weak detection 
even in a fourth beam), to a roughly rectangular region about 9 arcmin 
by 1 arcmin in size. ASAS data for this area again reveal no candidate short-period optical  
variables to 14~mag. This SMC field has  
also been monitored for four years by the Optical Gravitational Lensing 
Experiment (OGLE-IV) using the 1.3~m Warsaw University telescope in 
Chile \citep{udalski15}. Within the FRB error region, 
we find no short-period variable stars in OGLE-IV to $V=18.5$~mag.
%we at first suspected that
% OGLE-SMC734.10.47 ($I=17.2$~mag, $V-I$=1.2) is variable with a 
%$P=9.6$~hr period and low 0.05~mag amplitude, but then realized the variations
%are an observational alias artifact. 
%To further test this 
%possibility, we obtained
%a spectrum of the star with the Magellan-1 6.5~m telescope at Las Campanas
%with the IMACS Long Camera on 
%July 23, 2014. A long-slit 5-min exposure with the 600 line mm$^{-1}$ grating 
%covered the 3650--6750~\AA\ range with $\sim 1$~\AA~resolution. 
%Fitting the spectrum with stellar templates from both the 
%ELODIE \citep{prugniel01} and \citet{pickles98} 
%libraries, the spectrum is consistent main sequence stars in the range
%of spectral types from F8 to G5, 
%with the best fit being a G2V star, supporting the case for a W Uma system.
  From the OGLE-IV data in the SMC region \citep{pawlak13}, we have counted
the number of periodic variables redder than $V-I=0.5$ with periods less
than 1 day, to $V=18.5$~mag, and find a density of 0.5 deg$^{-2}$. These 
variables should include all main-sequence contact binaries, including 
W UMa types. Despite the color cut, this estimate might be contaminated
by reddened main-sequence B-type contact binaries in the SMC itself,
and therefore the density is a conservative upper limit. The product 
of the density and the error area gives a chance probability of 0.13\%
of finding a W UMa within the error region.

The field of FRB121102, discovered by \citet{spitler14} with Arecibo, 
is north of the $+28^{\circ}$ declination limit of ASAS, and
in the mid-plane of the Galaxy (latitude $b=-0.2^{\circ}$),
where crowding and extinction make the type of optical followup decribed here difficult.

\subsection{Association probability of FRBs and W~UMa's}
Table~\ref{frbtable} summarises some of the parameters of the ten known FRBs and 
the candidate stellar counterparts that we have found for some of them.
In view of the above results, we re-visit the question of the probability that
FRBs and W~UMa's are associated. In 8 FRB fields that we have searched,
we have found one or two possible associations with a W UMa system, as follows.
Within the beam of FRB110703, there is the W UMa system discovered by \citet{loeb14}, with 
a 0.5\% random probability for such a system within the FRB beam. However,
as already mentioned, to avoid {\it a posteriori} statistics, this field and 
this system perhaps should not be counted. In FRB110627, there is a W UMa
separated by 29~arcmin from the FRB beam center, but a side-lobe detection 
of a source at that position is a real possibility, given the receiver orientation at the time, 
with a 5\% estimated chance probability 
for a W UMa. Apart from these two W UMa stars, 
in the field of FRB010621 we have found a likely SPB pulsator.
We have not been able to show that it is connected to variability in a low-mass star,
and it is unclear whether or not the SPB star itself is related to the FRB. We will
therefore not consider it further in the accounting below.

As seen in Table 1, five of the eight FRB fields have an {\it a priori} random probability of 0.5 to 0.8\% of
hosting a W UMa system, two fields have a probability of 4--5\%, and one field 
(the SMC field, which is well localized) has a low probability 
of 0.13\%. To estimate the random probability of the experimental result, we find the binomial probabilities
of the observed number of successes, or more, in each of the three sets 
of fields with similar probabilities (for the 0.5-0.8\% fields
we conservatively assume 0.8\%). We then multiply the three binomial probabilities. 
The associations between FRBs and variable stars may exist in: zero out 
of 7 cases (if we do not count the first case, of FRB110703, and none of the other 
associations is real); in 
1 of 7 cases (if the FRB110628 association is correct, and then the random binomial probability is  9\%); or 2 of 8 cases 
(if we do count the case of FRB110703, in which case the random probability of the experimental result is $3.6\%$.) 
Finally, if our accounting includes all types of bright short-period
variable and potentially flaring stars, so that we include also the SPB star HD174628 in the field of
FRB010621, then the chance probability for the experimental result is
$(1.33-5)\times 10^{-3}$. 
Given this range, we cannot yet claim to have established an association between
the two phenomena. In any case, the fact that bright W UMa stars, or bright flaring stars
of any type, were {\it not} found in most of the fields, does not rule out the flare-star/FRB 
connection hypothesis. Stellar sources of FRBs will likely have a distribution of optical
luminosities and fluxes, possibly uncorrelated with the FRB luminosities, and therefore 
any flux-limited survey like ours for the optical counterparts will uncover only a fraction 
of the associations.      
 
\section{FRB110220 and FRB140514 are from the same repeating source}
\citet{petroff15a} re-monitored the sites of a number of FRBs with the Parkes Radio 
Telescope for a total of 85 hours on sky, leading to the real-time discovery of 
FRB140514 at the position of the previous FRB110220. Although only about 
21 hours out of the 85 
hours were spent on the field of FRB110220, all 85 hours were spent on {\it some} previous FRB 
location, and could have led to the discovery of a new FRB at the location of a previous one.
Therefore for the purpose of calculating the chance probability of detecting an unrelated 
FRB within a re-monitored FRB region, we must consider all 85 hours of on-sky time. 
\citet{petroff15a} estimated this chance probabilty at a high level of 32\%, 
and concluded that the
two bursts are from different sources. However, based on the available 
information and our own simple and straightforward calculation,
we contest this conclusion, as detailed below.

According to \citet{petroff15a}, the field of FRB110220 was observed 
with a cycle of five telescope pointings, called a 
``gridding'' pattern \citep{morris02}. In the
first pointing the central, 14-arcmin-diameter, beam of the receiver 
was pointed at the same position
as the center of the beam in which FRB110220 was detected. 
In the subsequent four pointings,
the telescope was pointed 9~arcmin north, south, east, and west of the first 
pointing, and then the cycle was repeated. The real-time FRB140514 was
detected only in the central beam of one 
of those 9-arcmin north-offset pointings.
(Actually, from comparison of the published coordinates, it appears the offset
was only about 5~arcmin north of the FRB110220 coordinates in \citet{thornton13}.)
At any given moment in the real-time survey, the central beam was the only one
covering part or all of the beam area of the original burst. 
Thus, for the purpose of estimating the 
chance probability of an unrelated burst occurring as close to the location of 
of the original burst as was, in fact, observed for the real-time burst, the 
monitored solid angle that needs to be considered is just the solid angle
covered by the central beam. 
%To err on the conservative side, we will assume
%that the real-time FRB would have been detected not only had it occurred
%within the 7-arcmin radius of the central beam, but out to the midpoint 
%between the central beam and the surrounding beams, i.e. about a 14 arcmin 
%radius from the center of the central beam. 
This solid angle is 
$\pi (7/60)^2$~deg$^{2}=0.042$~deg$^{2}$. For an assummed all-sky FRB rate of 
3300~day$^{-1}$ \citep{rane15}, and the 85~hr on-sky time of the 
\citet{petroff15a} survey, one then expects 0.0125 of an event within 
the central beam. However 
proper account must also be taken of the fact that, during the entire survey, 
an FRB was {\it not} detected in any of the other 12 beams of the receiver,
%Assuming again 0.17$~$deg$^{2}$ covered per beam, 
within which 
we would have expected 0.15 of an event
outside the central beam, with a complementary probability for 
non-detection of 0.85. The probability for an unrelated FRB being detected in 
the central beam of the real-time experiment, and only there, is therefore
$0.0125\times 0.85=1\%$.
 Conversely, we can state with $99\%$ confidence that the 
original FRB and the real-time FRB are from the same, repeating, source.   
We note that, if we use the somewhat lower FRB rate estimates, 2500~day$^{-1}$ 
\citet{keane15,macquart15}or 2000~day$^{-1}$\citep{burke-spolaor14}, 
the chance probability will go down correspondingly
and the confidence of our conclusion rises further.

The DMs of the original and the real-time FRBs were 945 and 562~pc~cm$^{-3}$, 
respectively, i.e. a difference of almost a factor of 2. If, as we have  argued,
the two FRBs are from the same source, such a change  in the intervening 
intergalactic free electron column density over the course of about 3 years
is impossible. This then rules out intergalactic dispersion as the origin
of the large DMs of FRBs. Conversely, in the flare-star scenario, there is
no particular reason why the plasma blanket around a flare star would not 
change between bursts separated by years. Such changes are even 
to be expected, given that the coronal plasma should be outflowing, as argued
by \citet{loeb14}. A repeating FRB in the same beam with changing DM 
could still arise in an extragalactic scenario, but only if the DM 
is produced by electrons local to the source (e.g. \citet{connor15}).
The source could be a true repeating one 
(i.e. the source is not destroyed in a burst),
or there could be several separate FRB sources in an ``FRB-active'' 
galaxy, or cluster of galaxies.   
   
\section{Theoretical considerations regarding the flare-star model}

We now address a number of arguments that have been raised against
the flare-star hypothesis of FRBs.
One claim has been that the coronal plasma blanket around a flaring star would
necessarily be too dense to produce the observed strict $f^{-2}$ frequency dependence of 
the pulse arrival times of FRBs \citep{dennison14,tuntsov14}. For the $f^{-2}$ behavior, 
a density $n_e<10^9$~cm$^{-3}$ is 
required since the plasma frequency is
$\nu_p=(\omega_p/2\pi)=0.9~{\rm GHz}(n_e/10^{10}{\rm cm^{-3}})^{1/2}$
\citep{rybicki86}. However, for example, \citet{getman08a,getman08b} have found that 
for the brightest X-ray flares
from pre-main-sequence stars, the coronas have electron 
column densities in the range $N_e=10^{21-22}$~cm$^{-2}$
(as implied by the DMs of FRBs) but also length scales as high as $R\sim 10^{12-13}$~cm, 
indicating mean densities of $n_e\sim 10^{8-9}$~cm$^{-3}$. As such, one would expect to 
see the $f^{-2}$ frequency sweep in stellar radio flares. Indeed, there are such examples.
\citet{osten08} show high-time-resolution radio data for the active young M-star 
AD Leonis, with a clear delay in signal arrival time with decreasing frequency. The typical drift 
rates are 2 GHz s$^{-1}$. Although, in this specific star, the observed drift rate is 
 an order of magnitude larger than seen in FRBs, and it is hard to say what is the exact 
frequency dependence of the sweep, it shows that known stellar flares
do sweep in radio frequency, at least roughly in the required sense. We encourage radio observers
of stellar flares to obtain more such high-time-resolution data of more stars, to broaden 
the known phenomenology of these effects.

Another argument against a stellar-flare origin for FRBs 
has been that radio emission would be suppressed through free-free absorption
in the corona that produced its DM if the corona is limited by the radius of 
the star, $R_\star$,
and its temperature is limited by the virial temperature of the star, $\sim 10^7$K \citep{luan14}. 
However, the optical depth for free-free absorption at a frequency of $1.4$ GHz 
declines with increasing temperature $T$ and size $R$ 
of the emitting region as \citep{kulkarni14},
\begin{equation}
\tau_{\rm ff}= 0.1 \left({T\over 10^8~{\rm K}}\right)^{-3/2} 
\left({R\over 10^{13}~{\rm cm}}\right)^{-1} 
\left({DM\over 10^3~{\rm cm^{-3}~ pc}}\right)^2,
\end{equation}
and becomes negligible for the observed DMs of FRBs 
at  $T>10^8$ K and $R\gg 10R_\star$. 
Observationally, \citet{getman08a} and \citet{getman08b} show 
that pre-main-sequence  flare
stars often have flares with
temperatures $T\sim 10^{8-9}$K
over a spatial scale $R\sim 10^2 R_\star\sim 10^{13}~{\rm cm}$, 
invalidating the argument for significant free-free absorption.
The free-free optical depth scales with frequency as $\nu^{-2}$,
and therefore for the above parameters of real stellar flares, we would
not expect to detect FRBs at frequencies much lower than 1.4~GHz. This 
is consistent with a recent strong limit on FRB detection at 145~MHz obtained by 
\citet{karastergiou15} using the LOFAR array, and requiring that FRBs have a rising 
flux density spectrum between these two frequencies.  
  
\citet{kulkarni14} have argued that radio emission from flare stars would 
necessarily be accompanied by bright X-ray flares. Considering
the rate of FRBs and comparing it to the known statistics of all-sky X-ray  
variability, they concluded that there is a shortage of known X-ray flares, and
hence FRBs cannot originate from flaring stars. Flaring stars indeed erupt in 
both radio and X-ray wavelengths. There is a well-known correlation, over 
many orders of magnitude in luminosity, between the radio and X-ray 
luminosities of flares from stars with active coronae, whether between the
peak luminosities or the time-averaged luminosities \cite[e.g.][]{benz94}. 
However, there is no one-to-one correspondence between individual 
radio and X-ray flares, as would be required in order to derive an expected
rate of X-ray flares from the observed FRB rate. For example, \citet{williams15} have monitored a 
close M-star binary simultaneously
in many bands, including radio and X-rays, and report a general lack of 
correlation between flares in any two bands.   

Thus, none of the above arguments against the flare-star origin of FRBs is valid, mainly
because the observed properties of some stellar flares already show that radio flares with 
characteristics similar to FRBs are possible. A recent addition to this list of similarities
is the high circular polarization measured in  FRB140514 \citep{petroff15a}; radio bursts from 
flaring stars are often 100\% circularly polarized.
  
Most recently, \citet{katz15} showed that the non-monotonic dependence of burst 
widths on dispersion measure excludes the intergalactic medium as the 
location of scattering that broadens the FRBs in time. This argues in favor of a 
local origin for FRBs, where the observed DM is intrinsic to the FRB sources,
as expected in the case of flaring Galactic stars. 

\section{Conclusions}
The ten published detections of 
FRBs have triggered a wealth of proposed interpretations, ranging from exotic processes at 
cosmological distances to atmospheric and terrestrial sources. 
\citet{loeb14} have previously suggested
that FRB sources could be nearby flare stars, and pointed out the presence 
of a W-UMa-type contact binary within the beam of one out of three FRB
fields that they examined. 
Using time-domain optical photometry and spectroscopy,
we now find a possible W UMa counterpart to another FRB, and
a rare slow-pulsating B star in the beam of a third FRB.  
The random probabilities for  having one or two real W UMa's 
within the eight studied FRB beam areas
are $\sim 9\%$ and 4\%, respectively.

Contrary to previous claims, we conclude with $99\%$ confidence
that  two FRBs which were discovered 3 years apart 
within the same radio beam are from the 
same repeating source. The different  DMs of the two bursts 
then rule out a cosmological origin for the DMs, but are consistent with the flare-star 
scenario with a varying plasma blanket between bursts. Finally,
we have shown that the theoretical objections that were raised 
against a Galactic origin of FRBs 
are incorrect because they are circumvented by the observed properties of 
some stellar radio flares \citep{getman08a,getman08b}.   
We conclude that the flare-star-origin hypothesis for FRBs is still a strong player 
in the game, and is consistent with all current observations. 
In contrast, the idea that the 
intergalactic medium is responsible for the DM of FRB sources is in 
direct contradiction to the observation of a changing DM between two FRBs that, as we have argued, 
are almost certainly from the same repeating source.
 
\section*{Acknowledgments}
We thank E. Ofek for his input regarding the localisation of the Lorimer burst,
%and T. Mazeh for advice. 
J. Miralda-Escude for comments,
W. Freedman and B. Madore for data
obtained at the Magellan Telecopes, and the anonymous referee for 
very useful suggestions.
A.L. acknowledges support from the Sackler Professorship
by Special Appointment at Tel Aviv University. This work was supported in part
by NSF grant AST-1312034 (A.L.) and Grant 1829/12 of the I-CORE program
of the PBC and the Israel Science Foundation (D.M. and T.M.).
The research by T.M. leading to these results has received funding from the European Research Council under 
the EU's Seventh Framework Programme (FP7/(2007-2013)/ERC Grant 
Agreement No.~291352), and Israel Science Foundation grant No.~1423/11 to T.M. 
Research by Y.S. is supported by an appointment to the 
NASA Postdoctoral Program at
the Jet Propulsion Laboratory, administered by Oak
Ridge Associated Universities through a contract with NASA.
R.M.R. acknowledges support from
NSF grant AST-1413755. Research at Lick Observatory is 
supported by the University of California and partially supported by 
a generous gift from Google. M.K. acknowledges support by the Polish National Science Center 
under grant  DEC-2011/03/B/ST9/03299.

\bibliography{wdbib} \bibliographystyle{apj}

%\onecolumn
%\newpage

\end{document}